\documentstyle[prb,preprint,aps,eqsecnum]{revtex}

\begin{document}
\draft
\preprint{preprint}
\title{
The Fermion Chern-Simons Gauge Theory of Fractional Quantum Hall Effect 
for  Electromagnetic Polarization Tensor} 
\author{ Tae-Hyoung Gimm\cite{gimm} and Sung-Ho Suck Salk\cite{salk} }
\address{ Department of Physics, 
Pohang University of Science and 
Technology, Pohang 790-784, Korea}
\date{\today}
\maketitle

\begin{abstract}
Unlike an earlier theory,
 by avoiding 
both the electromagnetic gauge field shift and the assumption of the zero average of electromagnetic field fluctuation
 the fermion Chern-Simons gauge theory is 
reformulated to obtain mean field solutions and a self-consistent 
expression of the electromagnetic polarization tensor 
in terms of the composite fermion picture
for the systems of fractional quantum Hall effect.
Thus the newly derived  electromagnetic polarization 
tensor is shown to depend 
on  the residual (effective) magnetic field `seen' by composite fermions
rather than the 
statistical field, which differs from the earlier
theory. The present theory reproduces the Hall conductance of fractional 
quantum Hall effect. 
The self-consistent picture of the composite 
fermion is maintained in all of our derivations:
both the mean field solutions and 
the electromagnetic  polarization tensor are described by the 
residual magnetic field seen by
the composite fermions.
\end{abstract}
\bigskip

\def\ba{\begin{array}}
\def\ea{\end{array}}
\def\bc{\begin{center}}
\def\ec{\end{center}}
\def\be{\begin{eqnarray}}
\def\ee{\end{eqnarray}}
\def\bi{\begin{itemize}}
\def\ei{\end{itemize}}
\def\bm{\left(\ba}
\def\em{\ea\right)}
\def\bn{\begin{eqnarray*}}
\def\en{\end{eqnarray*}}
\def\bt{\begin{tabular}}
\def\et{\end{tabular}}
\def\bdes{\begin{description}}
\def\edes{\end{description}}
\def\ds{\displaystyle}
\def\sss{\scriptscriptstyle}
\def\mc{\multicolumn}
\def\nc{\nonumber \\}
\def\p{{\bf p}}
\def\A{{\cal A}}
\def\vA{{\bf A}}
\def\va{{\bf a}}
\def\tA{\tilde{{\cal A}}}
\def\cE{{\cal E}}
\def\E{{\bf E}}
\def\vD{{\bf D}}
\def\vd{{\bf d}}
\def\B{{\cal B}}
\def\vB{{\bf B}}
\def\R{{\bf R}}
\def\J{{\bf J}}
\def\j{{\bf j}}
\def\x{{\bf x}}
\def\r{{\bf r}}
\def\k{{\bf k}}
\def\F{{\bf F}}
\def\Q{{\bf Q}}
\def\q{{\bf q}}
\def\H{{\cal H}}
\def\O{{\cal O}}
\def\L{{\cal L}}
\def\Z{{\cal Z}}
\def\S{{\cal S}}
\def\N{{\cal N}}
\def\l{{\bf l}}
\def\lg{\langle}
\def\rg{\rangle}
\def\la{\leftarrow}
\def\ra{\rightarrow}
\def\Ra{\Rightarrow}
\def\da{\downarrow}
\def\lra{\leftrightarrow}
\def\Ua{\uparrow}
\def\a{\alpha}
\def\b{\beta}
\def\d{\delta}
\def\e{\epsilon}
\def\f{\frac}
\def\g{\gamma}
\def\h{\hbar}
\def\k{\kappa}
\def\l{\lambda}
\def\n{\nabla}
\def\o{\omega}
\def\p{\partial}
\def\s{\sigma}
\def\t{\times}
\def\v{\varepsilon}
\def\D{\Delta}
\def\G{\Gamma}
\def\L{\Lambda}
\def\Si{\Sigma}
\def\O{\Omega}
\def\i{\infty}

\section{Introduction}
Recently the fermion Chern-Simons theory has been of great interest to study
the systems of fractional quantum Hall effect (FQHE).
Based on the composite fermion theory initiated by Jain,\cite{jain} the fermion Chern-Simons theory of 
the FQHE system is advanced by Lopez and Fradkin\cite{lopez,fradkin} and Halperin and coworkers.
\cite{halperin} 
Jain\cite{jain} proposed the transformation of interacting 2-D electrons under an external
magnetic field to
composite fermions.
To be more specific, 
he interpreted the Laughlin ground state\cite{laughlin} as a state in which the interacting electrons bind flux quanta to screen out part of the external magnetic field by fastening each electron with an even number of flux quanta and these composite fermions exactly fill an integer number of Landau levels.

Lopez and Fradkin\cite{lopez,fradkin} proposed the fermion Chern-Simons theory of the FQHE by allowing fermions coupled to the statistical gauge field  to form the composite fermions; 
the partition function is given by
\be \label{aa}\Z (A) = \int {\cal D}\psi^{\ast}{\cal D}\psi{\cal D} a_\mu\exp(i\S), \ee
where the action $S$ is explicitly,
\be \label{bb}\S&=&\int d^3z\left\{
\psi^{\ast}\left[i\frac{\partial }{\partial t} - e(A_0 - a_0) + \mu\right]\psi(z)\right.
-\frac{1}{2m}\left|[\nabla+ie({\bf A}-{\bf a})]
\psi(z)\right|^2+\\ \nonumber
& &+\left.\frac{\a}{2}\varepsilon^{\mu\nu\rho}a_{\mu}
\partial_{\nu}a_{\rho}\right\}
-\frac{1}{2}\int d^3z\int d^3z^{\prime}\left(|\psi(z)|^2-\bar{\rho}\right)
V(z-z^\prime)
\left(|\psi(z^\prime)|^2-\bar{\rho}\right),  \ee
Here $\psi(z)$ is a second quantized Fermi field and  $\mu$, the chemical
potential.
$A$ and $a$ are the electromagnetic vector potential and the statistical gauge potential respectively.
$V$ is the pair interaction and $\bar{\rho}$, the average electron density.
The resulting Chern-Simons `Maxwell' equation
of the $\mu = 0$ component is
\be \varepsilon^{ij}\partial_{i}a_{j}(x) = \frac{e}{\a}\rho(x) = b.
\ee
Here $b$ is the statistical `magnetic' field and $\rho(x)$, the electron density.
$\a$ is the Chern-Simons (statistical) coupling constant given by $\a = \frac{e}{n\phi_0}$ 
or $\a = \frac{1}{2\pi n}$ for $\hbar=e=c=1$ with the choice of an even number $n$ of the statistical flux quanta $\phi_0$.
The action above involves the dynamics of a system of spinless interacting fermions
coupled to both the electromagnetic and statistical (Chern-Simons) gauge fields.

In the present study we evaluate the effective action and present a generalized theory of
electromagnetic 
response
functions by avoiding the gauge shifts and
the assumption of the zero average of external electromagnetic 
fluctuation,
unlike the original theory of Lopez and Fradkin.\cite{lopez,fradkin}
Both the mean field solutions and the electromagnetic polarization function are described by the residual (effective)
magnetic field `felt' by the composite fermions.

\section{Mean Field Solutions and Electromagnetic Response Functions}
In the fermion Chern-Simons theory, the statistical
gauge field does not completely screen out the
electromagnetic gauge field. Thus there exists an residual (effective) gauge field, that is, the unscreened portion
of the electromagnetic gauge field, $A^{\rm eff}_{\mu}$,
\be \label{1} A^{\rm eff}_{\mu} \equiv A_\mu-a_\mu. \ee
The effective (residual) magnetic field is then 
\be \label{ab} \vec{B}^{\rm eff} = \vec{\nabla}\times\vec{A}^{\rm eff}. \ee
In the approach 
of Lopez and Fradkin,\cite{lopez,fradkin} a small fluctuating electromagnetic field is introduced into the Chern-Simons term by allowing gauge   
field shifts.
They also assumed the vanishing average of the external electromagnetic fluctuation for the evaluation of various mean field solutions.
In the following we avoid such gauge field shifts and the assumption of the vanishing average of the external electromagnetic fluctuation.

The Hubbard-Stratonovich transformation of
the quartic fermion field term in (\ref{aa}) and 
the integration over the fermi fields yields 
the effective action,\cite{lopez,fradkin}
\be \label{efas} \S_{\rm eff} = -i{\rm Tr} \ln\left(iD_0+\mu+\lambda- \frac{1}{2m}{\bf D}^2
\right) + \S_{\rm CS}\left[a_{\mu}\right] + \S\left[\lambda\right]
\ee
where  
\be \label{sl} \S [\lambda] =
-\int d^3z\lambda(z)\bar{\rho}
+\frac{1}{2}\int d^3 z\int d^3
z^\prime\lambda(z)V^{-1}(z-z^\prime)\lambda(z^\prime). \ee

In order to obtain the classical configurations $\bar{a}_\mu (z) \equiv \lg a_\mu (z)\rg$ and $\bar{\lambda} (z) \equiv \lg \lambda (z)\rg$, we introduce the semiclassical (saddle point) approximation by requiring that $S_{\rm eff}$ be stationary under small fluctuations; 
\be \label{seq} \left.\frac{\delta \S_{\rm eff}}{\delta a_{\mu}(z)}\right|_{\bar{a},\bar{\lambda}} = 0 
\hspace{0.5cm}{\rm ~and}
\hspace{0.5cm}\left.\frac{\delta \S_{\rm eff}}{\delta \lambda(z)}\right|_{\bar{a}, \bar{\l}} = 0.\ee
By introducing $a_\mu=A_\mu-A_\mu^{\rm eff}$ into 
the Chern-Simons term of (\ref{efas}), the equations of motion from 
the variation of $\S_{\rm eff}$ with respect to $a_\mu (z)$ and $\lambda (z)$ 
in (\ref{seq}) are obtained as
\be \label{4} \left\langle j_{\mu}(z)\right\rangle
 - \frac{1}{n\phi_0}\varepsilon^{\mu\nu\rho}
\{\left\langle \partial_{\nu}A_{\rho}(z)\right\rangle-\left\langle\partial_{\nu}A^{\rm eff}_{\rho}(z)
\right\rangle \} = 0, \ee
and
\be \label{l} \left\langle j_{0}\right\rangle - \bar{\rho}
+ \int d^3z^\prime V^{-1}(z-z^\prime)
\left\langle\lambda\left(z^\prime\right)\right\rangle =0. \ee
Here it is of note that the above equation of motion (\ref{4}) differs from
that of Lopez and Fradkin;\cite{lopez} the first term in their equation (3.10) in Ref. 2 is 
given by the difference between the statistical field strength tensor ${\cal F}^{\mu\nu}$
and the electromagnetic field strength tensor $F^{\mu\nu}$. 

The condition for the uniform liquid state is obtained from (\ref{l}), 
\be \label{2} \left\langle j_{0}\right\rangle = \bar{\rho}, \ee
for $\left\langle \lambda (z) \right\rangle = 0$. 
From (\ref{1}) and (\ref{ab}), the Chern-Simons (statistical) magnetic field is given by  
\be \label{3} \vec{b} = \vec{\nabla} \times \vec{a} = \vec{B} - \vec{B}^{\rm eff}. \ee
Using (\ref{2}) and (\ref{3}) for (\ref{4}) and setting $\mu=0$,
we find
the uniform average statistical magnetic field,
\be \label{babo} \left\langle b\right\rangle = n\phi_0\bar{\rho}. \ee
From (\ref{3}) and (\ref{babo}) the effective (remaining) external magnetic field is then
\be \label{6} B^{\rm eff} = B - \left\langle b\right\rangle = B - n\phi_0\bar{\rho}. \ee
In coincidence with the Jain's composite fermion picture, each electron is 
now assumed to be dressed by an even number of statistical (Chern-Simons) flux 
quanta, $n=2m$ with $m$, an integer.
The number of effective (residual) flux quanta, $N^{\rm eff}_{\phi}$ is then from (\ref{6}) above,
\be \label{7} N^{\rm eff}_{\phi} = N_{\phi} - 2mN_e.  \ee
Here $N_\phi$ is the total number of applied (external) flux quanta and $N_e$, the total number of electrons.
From the filling fraction defined by  $\nu =N_e/N_\phi$ and the relation (\ref{7}), we readily obtain the fractional filling factor,
\be\nu = \frac{p}{2mp + 1}\ee
with  $p = \frac{N_e}{N^{\rm eff}_{\phi}},$
and the effective Landau gap,
\be \hbar \o^{\rm eff}_{c} = \frac{\hbar \o_c}{2mp+1}.\ee
As shown above, we were able to avoid the gauge field shifts and the assumption of the
vanishing average of fluctuating electromagnetic field in order to reproduce various mean field
solutions: the uniform liquid state, the fractional filling fraction, and the effective Landau gap.

In order to obtain the linear response functions, we now consider the Gaussian fluctuations around the uniform classical 
values;
$\d A^{\rm eff}_{\mu} (z)$ and $\d\lambda(x)$ are taken to be small
fluctuations around the mean (classical) values of the effective (residual) field $A^{\rm eff}$ and the collective mode $\lambda $, that is, 
$A^{\rm eff}_{\mu} \ra \lg A^{\rm eff}_{\mu}\rg + \d A^{\rm eff}_{\mu}$ and
$\l \ra  \lg \l \rg +\d\l$. 
In this  spirit of the linear response theory the effective action to quadratic order is written,
\be 
\label{seff}
\S^{(2)} &=&
\frac{1}{2}\int d^3xd^3y \delta A^{\rm eff}_{\mu}(x) \Pi_{\mu\nu}(x,y)
\delta A^{\rm eff}_{\nu}(y)+  \nonumber \\
 & &+\S_{\rm CS}\left[A_{\mu}-A^{\rm eff}_{\mu}-\delta_{\mu0}\d\lambda\right]
+\f{1}{2} \int d^3 z d^3 z'\d\l(z) V^{-1}(z-z') \d\l(z'),
\ee
after the shift of the scalar bose field is made.
The above expression is different from that of Lopez and Fradkin\cite{lopez,fradkin} in that 
$\Pi_{\mu\nu}$ is the polarization tensor associated with the residual gauge field, but not with the statistical gauge field.
The effective action involving the quadratic term above is then equivalent to the one-loop 
bubble contributions associated with the effective field $A^{\rm eff}_{\mu}$.

The expansion of the Chern-Simons term around $\lg \lambda \rg = 0$ yields
\be 
\label{sef1}
\S_{\rm CS}[A_{\mu}-A^{\rm eff}_{\mu}-\delta_{\mu 0}\d\lambda] &=& 
\S_{\rm CS}[A_{\mu}-A^{\rm eff}_{\mu}]+ \nonumber \\
& &-\frac{\alpha}{2}\int d^3z\d\lambda(z)
\left(B(z)-B^{\rm eff}(z)\right)+(h. o.) \\
&=& \S_{cs} [A_{\mu} - A^{\rm eff}_{\mu}] - \frac{\alpha}{2}\int d^3 z \d\lambda(z) b(z)+(h. o.). \nonumber
\ee
A scalar (collective) bose field term is, from (\ref{seff}) and (\ref{sef1}),
\be S^{(2)}[\d\l]
=\frac{\alpha}{2}\int d^3z\d\lambda(z) b(z) + \f{1}{2} \int d^3 z d^3 z'\d\lambda(z) V^{-1}(z-z')
\d\l (z').
\ee
After the Gaussian integration over the collective bose field $\lambda$, the effective action (\ref{seff}) becomes
\be \label{salk2} \S^{(2)}_{\rm eff} &=&
\frac{1}{2}\int d^3xd^3y \delta A^{\rm eff}_{\mu}(x)\Pi_{\mu\nu}(x,y)
\delta A^{\rm eff}_{\nu}(y)+  \nonumber \\
 & &+\S_{\rm SC}[A_{\mu}-A^{\rm eff}_{\mu}]+ \\
 & &-\frac{\alpha^2}{2}\int d^3z\int d^3z^\prime
b(z)V(z-z^\prime)b(z^\prime). \nonumber
\ee 
The polarization tensor in (\ref{salk2}) can be expressed in the low energy limit 
in terms of three gauge invariant tensors   
$E_{\rm eff}^2, B_{\rm eff}^2$ and Chern-Simons terms 
which are associated with the 
effective gauge field $A^{\rm eff}$, not with the statistical gauge field. Thus 
the coefficients of these tensors are namely the dielectric constant $\epsilon^0$, diamagnetic susceptibility $\chi^0$, and Hall conductance $\sigma^0_{xy}$  
in association with the response of the system of the composite fermions to the weak perturbation of the
residual field, $\d A^{\rm eff}_\mu$.

In order to integrate out the statistical gauge field, we rewrite (\ref{salk2}),
\be \label{salk3} 
\S^{(2)}_{\rm eff} &=&
\frac{1}{2}\int d^3xd^3y \delta(A_\mu(x)-a_\mu(x))\Pi_{\mu\nu}(x,y)
\delta(A_\nu(y)-a_\nu(y))+  \nonumber \\
 & &+\S_{\rm CS}[a_\mu]+ \f{1}{2\b}(\p_\mu a^\mu)^2 + \\
& &-\frac{\alpha^2}{2}\int d^3z\int d^3z^\prime 
b(z)V(z-z^\prime)b(z^\prime), \nonumber
\ee
where we introduced a gauge fixing term, $\frac{1}{2\beta}(\p_\mu a^\mu)^2$ in order to
avoid singularity in the inverse matrix of the  quadratic term in $a_\mu$ in momentum space.

We obtain from (\ref{salk3}) 
the following matrix for the quadratic term in $a_\mu$ in momentum space representation

\scriptsize
\be
M=\left[\ba{lll}\mbox{\bf q}^2\Pi_0 + \frac{\o^2}{\b}& \o q_1 \Pi_0 + i q_2 (\Pi_1 + \alpha)- \frac{\o  q_1}{\b}& \o q_2 \Pi_0 -i q_1 (\Pi_1 + \alpha)-\f{\o q_2}{\b}\\
\o q_1 \Pi_0 - i q_2 (\Pi+ \alpha)-\frac{\o q_1}{\b}& \o^2 \Pi_0 + q_2^2(\Pi_2 - \alpha^2 V)+\frac{q^2_1}{\beta} & - q_1 q_2 (\Pi_2 -\alpha^2 V)-i\o(\Pi_1+\alpha) +\frac{ q_1 q_2}{\b}\\
\o q_2 \Pi_0 + i q_1(\Pi_1 +\alpha)-\frac{\o q_2}{\b}&- q_1 q_2(\Pi_2 -\alpha^2 V) + i\o(\Pi_1 +\alpha)+\frac{q_1 q_2}{\b}& \o^2\Pi_0 + q^{2}_{1}(\Pi_2 - \alpha^2 V)+\f{ q_2^2}{\b}\\
\ea\right]
\ee
\normalsize
The determinant of the matrix above is then
\be
{\rm det} M = \frac{D(\omega, {\bf q})}{\b}(\omega^2+{\bf q}^2)^2, \ee
where
\be \label{D} D(\omega, {\bf q}) \equiv \Pi_0^2\omega^2-(\Pi_1
+\alpha)^2
+\Pi_0(\Pi_2-\alpha^2V({\bf q}^2)){\bf q}^2 \ee
with $\Pi_l=\Pi_l(\omega, {\bf q})$ with $l=0,1,2$.

The Gaussian integral can now be performed by taking the inverse
of the above $3\times 3$ matrix $M$. 
After the Gaussian integration over the statistical gauge field, the effective action for the electromagnetic fluctuation is  
\be \label{dd} \S^{\rm EM}_{\rm eff} = \frac{1}{2}\int d^3x
\int d^3y \delta A_\mu(x)K^{\mu\nu}(x,y)\delta A_\nu(y). \ee
Here the components of the above electromagnetic polarization tensor $K^{\mu\nu}$
in momentum space are obtained as 
\be \label{SS} K_{00} &=& {\bf q}^2K_0(\omega,{\bf q}),  \nonumber \\
K_{0j} &=& \o q_jK_0 (\o, {\bf q})+i\epsilon_{jk} q_kK_1(\omega,{\bf q}), \ \nonumber \\
K_{j0} &=& \o q_jK_0(\omega,{\bf q})-i\epsilon_{jk} q_kK_1(\omega,{\bf q}), \\
K_{ij} &=& \o^2\delta_{ij}K_0(\omega,{\bf q})-i\epsilon_{ij}\omega K_1(\omega,{\bf q})
+({\bf q}^2\delta_{ij}-q_iq_j)K_2(\omega,{\bf q}),\nonumber \ee
where 
\be \label{k1} K_0(\omega,\q) &=& -\frac{\alpha^2\Pi_0}{D(\omega,\q)}, \nc
K_1(\omega,\q)&=& \alpha+\alpha^2\frac{(\alpha+\Pi_1)}{D(\omega,\q)}
+\alpha^3V(q)\q^2\frac{\Pi_0}{D(\omega,\q)},\\
K_2(\omega,\q)&=& 
-\alpha^2\frac{\Pi_2+V(\q)(\omega^2\Pi_0^2-\Pi_1^2+\q^2
\Pi_0\Pi_2)}{D(\omega,\q)}. \nonumber
\ee
In the limit of small $\omega$ and ${\bf q}$, 
$\Pi_l(\omega, {\bf q})$'s with $l=0,1,$ and $2$ are obtained as 
\be
\label{pi}
\Pi_0(0,0) &=& \frac{pM}{2\pi B^{\rm eff}}=\frac{pM}{2\pi (B-b)}, \nonumber \\
\Pi_1 (0,0) &=& \frac{p}{2\pi},   \\
\Pi_2 (0,0) &=& 
-\f{p}{2\pi M} \nonumber \ee
with $M$ is the electron mass. 
$K_1$ in (\ref{k1}) leads to
\be \label{74} \lim_{{\bf q}\rightarrow 0}\lim_{\omega\rightarrow 0}K_1(\omega,{\bf q}) =
\frac{\alpha\Pi_1(0,0)}{\alpha+\Pi_1(0,0)}. \ee
Using (\ref{pi}) and $\alpha=\frac{1}{2\pi n}$ with $n=2m$ in (\ref{74}) above,
we reproduce 
the Hall conductivity 
\be K_1 (0,0)= \s_{xy} = \nu.\ee
which is a fractional multiple of $\f{e^2}{\hbar}$.

As shown in (\ref{pi}) above 
$\Pi_0(0,0)$ is seen to be different from the result of Lopez and Fradkin, that is, 
$\Pi_0(0,0) =\frac{pM}{2\pi b_{\rm eff}}$ with $b_{\rm eff}$ defined as
$a_1(x)=-b_{\rm eff}x_2$ and $a_2=0$ in the Landau gauge (or $\A_1(x)=-\B_{\rm eff}x_2$ and
$\A_2(x)=0$ using their notation for the statistical gauge potential $\A_\mu(=a_\mu)$ as 
shown in eq. (B3) in Ref. 2).
Thus unlike the theory  of Lopez and Fradkin,\cite{fradkin} the electromagnetic polarization tensor 
$K^{\mu\nu}$ is described by the effective magnetic field 
$B^{\rm eff}$ seen by the composite fermions
but not by the statistical  magnetic field $b$.

\section{Conclusion}
In the present study we treated the fermion Chern-Simons theory of FQHE
in a straightforward manner, by avoiding the gauge
field shifts. Further the assumption 
of the zero average of electromagnetic field fluctuations was not
necessary to reproduce various 
mean field solutions: the uniform liquid state,
the effective filling fraction, and the effective Landau gap.
Unlike the theory of Lopez and Fradkin,\cite{lopez,fradkin} the derived 
electromagnetic polarization tensor is described by 
the residual (effective) magnetic field rather than 
the statistical field, 
thus maintaining the self-consistent picture of the composite fermion. 
The present approach also reproduces 
the Hall conductance of 
the fractional quantum Hall effect. 
In all of our derivations the self-consistency 
of the composite fermion picture was maintained:
both the mean field solutions and the electromagnetic polarization tensor were
shown to depend on the residual magnetic field seen by the composite fermions.

\section{Acknowledgement}
\hspace*{0.4cm}One of the authors (SHSS) acknowledges the generous supports of 
Korean Ministry of Education under the BSRI program and the Center 
for Molecular Sciences at Korea Advanced Institute of Science and Technology.
He is also grateful to Pohang University of Science and Technology (Program P95002) for a 
partial financial support.

\end{document}